\documentclass{aa}        
\usepackage{txfonts}
\usepackage{graphicx,longtable,lscape,natbib,psfig,textcomp,fontenc,amssymb}  
\bibpunct{(}{)}{;}{a}{}{,} 
\newcommand{\ltsima} {$\; \buildrel < \over \sim \;$}  
\newcommand{\gtsima} {$\; \buildrel > \over \sim \;$}  
\newcommand{\lta} {\lower.5ex\hbox{\ltsima}}  
\newcommand{\gta} {\lower.5ex\hbox{\gtsima}}  
  
\newcommand{\Hb} {H$\beta$}  

\newcommand{\ergs}{\>{\rm erg}\,{\rm s}^{-1}}

\newcommand{\loiii}{L$_{\rm{\tiny{ [O~III]}}}$}

\newcommand{\forb}[2]{\mbox{$[{\rm #1\, #2}]$}}
\newcommand{\oiii}{\forb{O}{III}}

\begin{document}

\title{A pilot study of the radio-emitting AGN population: \\the emerging new
  class of FR~0 radio-galaxies}

\subtitle{} \titlerunning{The radio properties of FR~0 radio galaxies: a pilot
study} 
\authorrunning{Baldi, Capetti \& Giovannini}

\author{Ranieri~D. Baldi\inst{1,2,3}
\and Alessandro Capetti\inst{4}
\and Gabriele Giovannini\inst{5,6}}

\institute{SISSA-ISAS, via Bonomea 265, I-34136 Trieste, Italy \and
  Physics Department, The Technion, 32000, Haifa, Israel; \email{baldi@ph.technion.ac.il} \and Physics
  Department, Faculty of Natural Sciences, University of Haifa, 31905,
  Haifa, Israel \and INAF - Osservatorio Astrofisico di
  Torino, Strada Osservatorio 20, I-10025 Pino Torinese, Italy \and Dipartimento di Fisica e Astronomia, Universit\`a di Bologna, via
  Ranzani 1, 40127 Bologna, Italy \and INAF-Istituto di Radio Astronomia, via P. Gobetti
  101, I-40129 Bologna, Italy}


\date{}

\abstract{We present the results of a pilot JVLA project aimed at studying the
  bulk of the radio-emitting AGN population, unveiled by the NVSS/FIRST and
  SDSS surveys. The key questions are related to the origin of their
  radio-emission and to its connection with the properties of their hosts.

  We obtained A-array observations at the JVLA at 1.4, 4.5, and 7.5 GHz for 12
  sources, a small but representative sub-sample. The radio maps reveal
  compact unresolved or slightly resolved radio structures on a scale of 1-3
  kpc, with only one exception of a hybrid FR~I/FR~II source extended over
  $\sim$40 kpc. Thanks to the new high-resolution maps or to the radio
  spectra, we isolate the radio core component in most of them.

The sample splits into two groups. Four sources have small black hole (BH)
masses (mostly $\sim$10$^{7}$ M$_{\odot}$) and are hosted by blue
galaxies, often showing evidence of a contamination from star formation to
their radio emission and associated with radio-quiet (RQ) AGN. The second
group consists in seven radio-loud (RL) AGN, which live in red massive
($\sim10^{11}$ M$_{\odot}$) early-type galaxies, with large BH masses
($\gtrsim$10$^{8}$ M$_{\odot}$), and spectroscopically classified as Low
Excitation Galaxies (LEG), all characteristics typical
of FR~I radio galaxies. They also lie on the correlation between radio core
power and [O~III] line luminosity defined by FR~Is. However, they are more
core dominated (by a factor of $\sim$30) than FR~Is and show a deficit of
extended radio emission. We dub these sources 'FR~0' to emphasize their lack
of prominent extended radio emission, the single distinguishing feature with
respect to FR~Is. 

The differences in radio properties between FR~0s and FR~Is might be
ascribed to an evolutionary effect, with the FR~0 sources undergoing
to rapid intermittency that prevents the growth of large scale
structures. However, this contrasts with the scenario in which low
luminosity radio-galaxies are fed by continuous accretion of gas from
their hot coronae. In our preferred scenario the lack of extended
radio emission in FR~0 is due to their smaller jet Lorentz $\Gamma$
factor with respect to FR~Is. The slower jets in FR~0s are more
subject to instabilities and entrainment, causing their premature
disruption.

}

\maketitle

\section{Introduction}
\label{intro}

The advent of large area surveys with high sensitivity and the cross-match of
radio and optical data opened the possibility to use large samples of
extragalactic sources to investigate the links between the radio properties,
the central engine, and the host galaxies. Based on the recent results of
\citet{baldi09} and \citet{baldi10a}, a new population of radio-emitting
sources turns out to dominate in the local Universe.

\begin{figure*}
\includegraphics[scale=1.0]{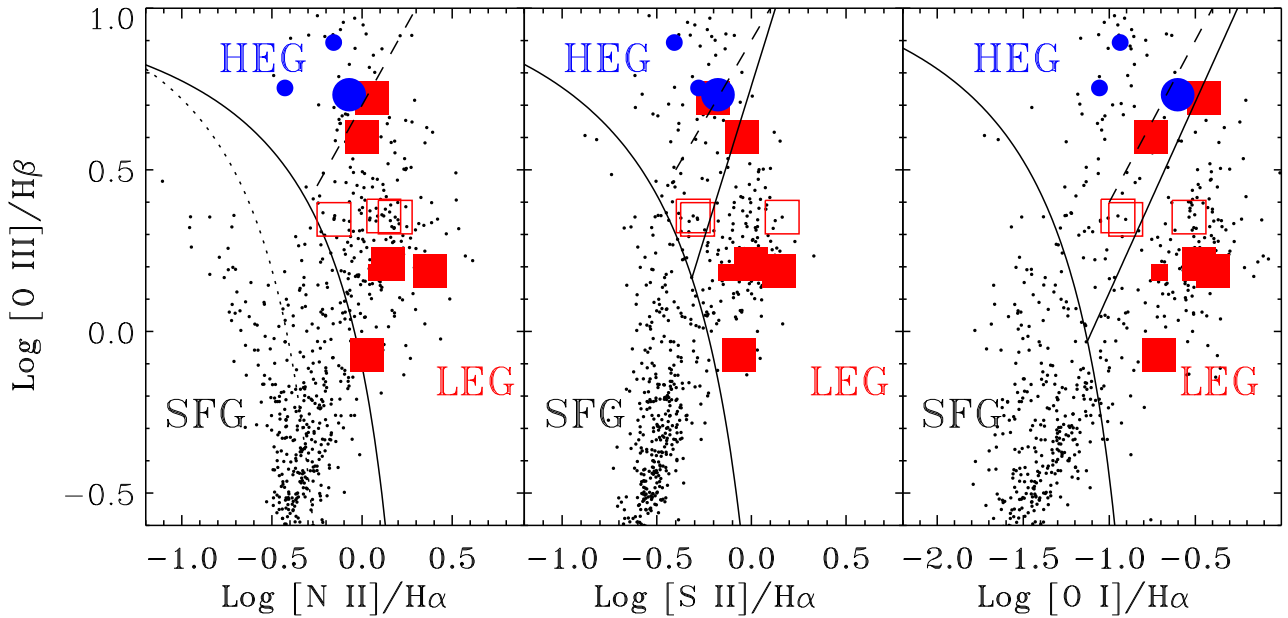}
\caption{Spectroscopic diagnostic diagrams for the galaxies of the
  sample.  The small dot points correspond to the SDSS/NVSS sample
  studied by \citet{baldi10a}, while the symbols in color represent
  galaxies with the new JVLA observations. The solid lines are
    from \citet{kewley06} and separate star-forming galaxies (SFG),
    LINER, and Seyfert. In the first panel the region between the two
    curves is populated by the composite galaxies. The dashed lines
    instead separate RL AGN into HEG from LEG \citep{buttiglione10}
    which roughly correspond to Seyfert and LINER classification for
    RL AGN. The red squares are the LEGs, while the blue circles are
  the HEGs. The large filled symbols are associated with BH masses
  $>$10$^{8}$ M$_{\odot}$, large empty with $\sim$10$^{7.7}$
  M$_{\odot}$, and small symbols with $<$10$^{7.3}$ M$_{\odot}$. }
\label{diag}
\end{figure*}

\citet{best05a} built a sample of 2215 radio-galaxies (RGs) by
cross-correlating the SDSS (DR2), NVSS, and FIRST datasets (hereafter the
SDSS/NVSS sample). This sample is selected at $F_{\rm 1.4} > 5$ mJy and it
includes RGs up to $z\sim$ 0.3, covering the range of radio
luminosity $L_{\rm 1.4} \sim 10^{22} - 10^{26}$ W Hz$^{-1}$. All morphologies
are represented, including twin-jets and core-jet FR Is, narrow and wide angle
tails, and FR IIs. However, most of them ($\sim$ 80 \%) are unresolved or
barely resolved at the 5$\arcsec$ FIRST resolution.

\begin{figure*}
\includegraphics[scale=0.75]{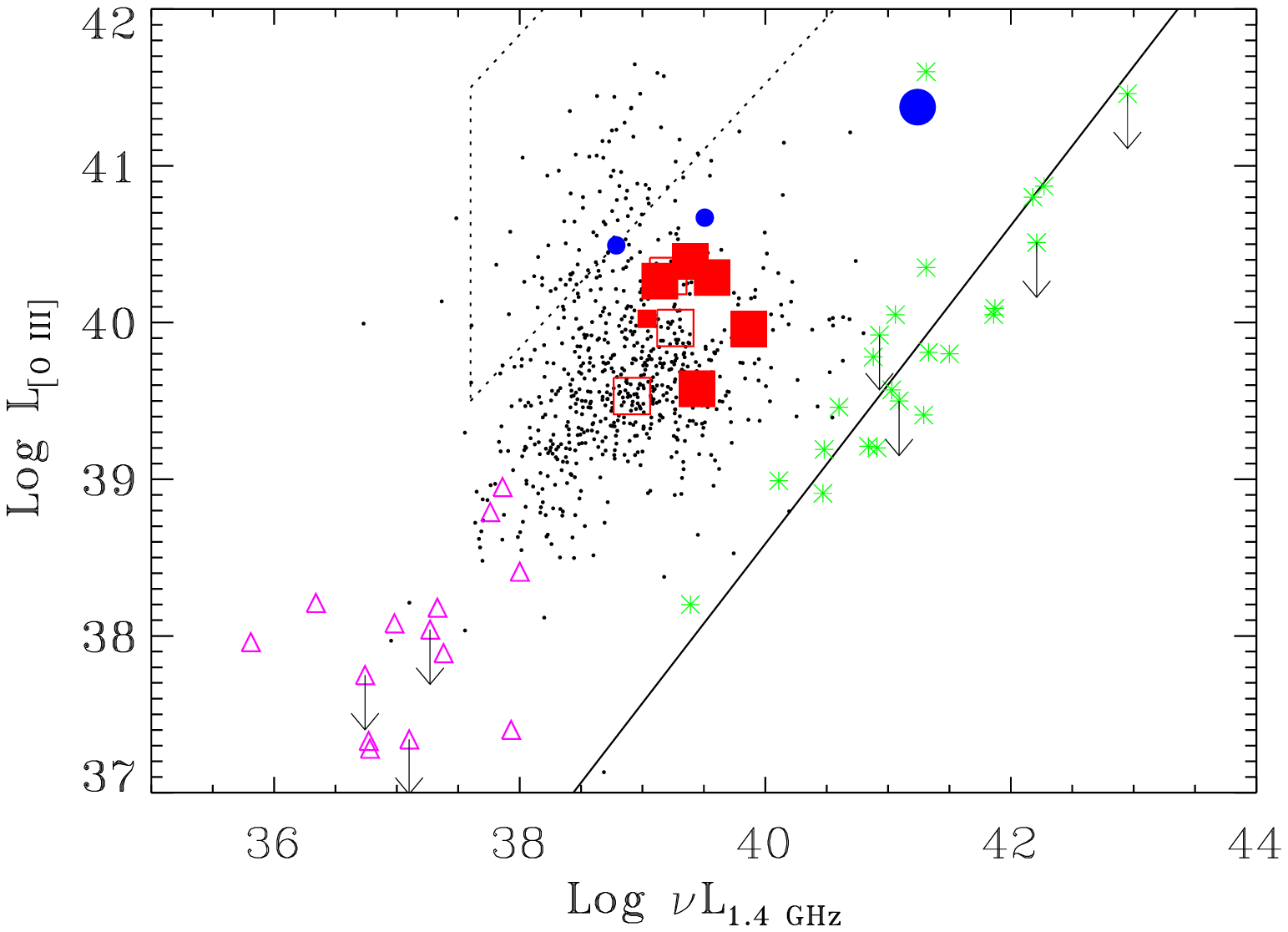}
\caption{FIRST vs [O~III] line luminosity (erg s$^{-1}$). The colored
  points represent the 12 galaxies observed with the JVLA coded as in
  Fig.~\ref{diag}.  The small dot points correspond to the SDSS/NVSS sample
  studied by \citet{baldi10a} . The solid line represents the correlation
  between line and radio-luminosity derived for the 3CR/FR~I sample. The
  dotted lines include the region where Seyfert galaxies are found.  The empty
  pink triangles are the CoreG and the green stars are the 3CR/FRIs.}
\label{lrlo3}
\end{figure*}

The SDSS/NVSS sample is mostly ($\sim$90\%) composed of RL AGN, representing
the bulk of the RGs population (with a space density $>$ 100 times higher than
3C sources) but it is virtually unexplored. A few partial results have been
already obtained, even though the radio data available are extremely limited,
generally only FIRST and NVSS images with angular resolution problems and the
missing radio spectral coverage (they are generally too faint to be detected
in the existing high frequency surveys). In a previous paper \citet{baldi10a}
found the following results on this sample:

1) the AGN power estimated from the optical line luminosity of the SDSS/NVSS
sample is at the same level of classical FR~I RGs. However, the
sample shows a deficit of a factor $\sim$100 of radio emission with respect to the
3C sources with the same optical power.

2) the SDSS/NVSS sample generally shows optical emission line ratios typical
of AGN: most of them ($\sim 70 \%$) can be classified as Low Excitation
Galaxies (LEG) similarly to the FR Is, but a substantial fraction of High
Excitation Galaxies (HEG) is also present\footnote{We bear in mind that HEG
  galaxies in the 3C catalog are all FR II RGs \citep{buttiglione10}}.

3) Most of the hosts of the SDSS/NVSS are statistically indistinguishable from
those of 3C sources from the point of view of morphology, color, stellar and
black hole masses. They are generally associated with giant elliptical
galaxies, with no signs of star formation, located in dense environment. Thus
the deficit in radio emission cannot be ascribed to differences in their
hosts. 

4) A small fraction ($\sim$10\%) of the sample shows rather different
photometric properties and it represents the RQ AGN contamination to the main
population of RL AGN.

\smallskip 
These results have been recently confirmed by the analysis of a larger sample,
obtained from the SDSS DR7, by \citet{best12}.

Similar results to the analysis of the SDSS/NVSS sample come from a parallel
study of a sample of 14 nearby giant early-type galaxies (ETGs) of extremely
low radio luminosity (10$^{20-22}$ W Hz$^{-1}$ at 1.4 GHz)
\citep{baldi09}. They are named CoreG and are much fainter than FR~Is
\citep{balmaverde06core}. Despite the low radio power, the CoreG host genuine
``miniature'' RL nuclei, indistinguishable from the FR~Is. In fact they extend
the FR~I relations at different wavelengths
\citep{chiaberge99,balmaverde06a,balmaverde06core,baldi10b}, indicating a
common synchrotron origin of the nuclear emission. However, CoreG show a less
extended radio structure and a core dominance up to $\sim$100 times higher
than FR~Is, but similar to the SDSS/NVSS sample.  A high core dominance is
generally interpreted as evidence of Doppler boosting in a radio source
oriented at a small angle with respect to the line of sight. The correlation
found with the core radio power and emission line luminosity (independent of
orientation) indicates that we are seeing a genuine deficit of extended radio
emission and that no geometric effect is present.  

However, the differences between CoreG and FR~Is might be driven by their much
lower radio luminosity.  For this reason a high resolution radio study of the
SDSS/NVSS sample, objects sharing the same range of power of FR~Is, is
required. In fact, the radio properties of the bulk of the RL population are
virtually unexplored and we have little or no information on their radio
morphology or on their spectral properties. This severely hampers our ability
to understand their nature and how they are related, on the one hand, to
classical RGs and, on the other hand, to low luminosity AGN. An
improvement of the radio data is needed to provide us with a more complete
view of the radio emission phenomenon.

We then undertook a project aimed at understanding the nature of the radio
emission of such objects and its connection with the host galaxies. Precisely,
we present new high-resolution observations with the JVLA of a pilot sample of
12 sources selected from the SDSS/NVSS sample. We obtained A array
observations in L and C bands, reaching a resolution of 0\farcs2. In
particular, the C band provides us with the possibility of measuring the high
frequency core emission, crucial for our purposes.

We aim at answering the following questions:

1) which is the nature of the radio emission: star formation, radio outflows as
in RQ AGN or genuine emission from a relativistic jet as in RL AGN? 

2) once we isolate the RL AGN, which is their level of core dominance?  Are
they highly core dominated similarly to the CoreG, but unlike the 3C sources?
Or, alternatively, they are just compact, but with well developed extended
radio structures which produce most of the radio emission?

3) do they follow the relation between radio core and line emission defined by
3C/FR~Is? The new observations aim at  isolating their genuine radio core emission,
an essential information to test their nuclear similarity with the FR~Is.

4) is their radio structure characterized by jet(s), diffuse plumes or
a double-lobed morphology? Are these jets one or two sided? The
distribution of jet sidedness will constrain their speed. 

5) which is the fraction of Compact Steep Spectrum in the sample? 

6) is there a relationship between the radio properties and the
optical spectroscopic classification, i.e., HEG and LEG differ from the
point of view of their radio structure? 

The paper is organized as follows. In Sect. \ref{sample} we define the
sample. Section~\ref{obs} presents the new JVLA observations for 12
sources.  In Sect.~\ref{results} we analyze the radio and
spectro-photometric properties of the sample. We discuss the results
in Sect.~\ref{discussion}, presenting a new class of radio-sources,
FR~0s (Sect.~\ref{FR0}). The summary and conclusions to our findings
are given in Sect.~\ref{conclusion}. We provide the notes on the radio
properties of the sources which show extended morphology in the
Appendix A.

\section{The sample}
\label{sample}
The objects observed with the JVLA have been extracted from the
SDSS/NVSS \citeauthor{best05a} sample adopting the following criteria: 

\begin{itemize}

\item redshift $z<0.1$,

\item the main optical emission lines detected at least 5 $\sigma$ significance
(for the \Hb\ we relaxed this requirement to 3 $\sigma$) to ensure a reliable
spectral classification, 

\item equivalent width for the \oiii\ lines larger than
3 \AA\ and with a measurement error smaller than 1 \AA; this ensures that the
line emission is associated with an AGN and not with a stellar origin \citep{capetti10}, 

\item declination in the range $-10 < \delta < 10$ in order to optimize the
observing strategy.

\end{itemize}

This procedure returns 68 objects. Twelve of them have been then randomly
selected for the observations.  In Tab.~\ref{table1} we give the main
properties of the observed objects.

We use the spectroscopic diagnostic diagrams
defined by \citet{buttiglione10} for the 3CR sample to recognize the nature of
their nuclear emission. These diagnostics are formed by pairs of nuclear
emission line ratios, to separate active nuclei from star-forming galaxies
(e.g. \citealt{baldwin81}), and, furthermore to differentiate AGN into
branches of different excitation level, i.e. low-excitation galaxies (LEG) and
high-excitation galaxies (HEG). This separation is similar to that introduced
by \citet{kewley06} as diagnostic to separate among the RQ AGN population,
LINER and Seyfert \citep{heckman80}.

From the point of view of their optical spectral classification, our sample
covers the whole range of ionization (Fig.~\ref{diag}), with eight objects
located into the LEG area and two objects falling into in the HEG region. Two
objects (ID~590 and 625) straddle the separation between the two
classes; we classify 590 (625) as LEG (HEG) since in 2 out of three
diagnostic ratios it is consistent with such a classification.

In Fig.~\ref{lrlo3} we show their location in a diagram comparing the FIRST
radio and optical [O~III]. All objects show a large deficit of radio emission
when compared to the FR~Is part of the 3C sample by a factor ranging from
$\sim$ 30 to $\sim$ 1000. They are all (but one) located in the region of
higher density of the SDSS/NVSS sample, just avoiding the lowest line
luminosity, due to the third requirement for their selection. ID~656 is on the
boundary of the region of RQ Seyferts.  Only one object (ID~625) falls in a
poorly populated area, being one of the objects of higher radio and line
luminosity. In Fig.~\ref{lrlo3} we  also stress the similarity of our
sample to the CoreG sample in terms of L$_{FIRST}$/\loiii\ ratio.

\begin{figure}
\includegraphics[scale=0.40]{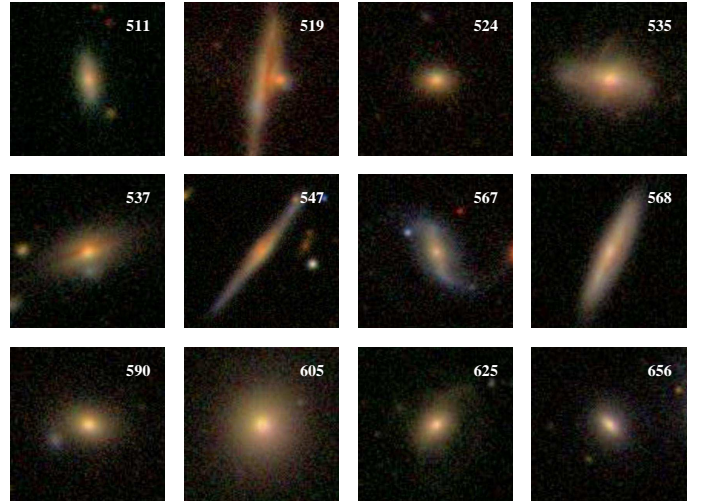}
\caption{SDSS optical images of the galaxies of our sample. The size of each
  frame is 40\arcsec $\times$ 40\arcsec.}
\label{SDSSimage}
\end{figure}

Fig.~\ref{SDSSimage} shows the SDSS optical images of the galaxies of
the sample. Most of the targets has an optical morphology akin of
elliptical galaxies. Some, namely ID~519, 537, 547 and 568, show
edge-on disks, while 567 shows spiral arms. 

\begin{table}
  \caption{The sample}
\begin{tabular}{l|c|c|c|c|c|c|c}
  \hline
  ID  & R.A. & Dec.& z & F$_{\rm NVSS}$ & F$_{\rm FIRST}$ & F$_{\rm [O~IIII]}$ \\
  \hline
  511 &   353.1122 &-1.2017 &0.094 &  3.2 & 1.57 &  92 \\
  519 &   354.0168 & 0.0798 &0.076 & 18.6 & 16.53 & 138\\
  524 &   356.5379 & 0.9858 &0.093 & 24.8 & 25.92 &  43\\
  535 &   359.4337 &-0.1749 &0.076 &  7.2 &  4.70 & 133\\
  537 &     0.4175 & 1.0920 &0.061 &  6.5 &  6.49 &  38\\
  547 &     5.1446 &-0.4708 &0.072 & 10.7 &  9.53 &  75\\
  567 &     9.9076 &-0.3288 &0.055 & $<$8.3 &  2.86 & 148\\
  568 &     8.6813 &-0.0408 &0.042 & 57.0 & 56.01 &1161\\
  590 &    15.2546 &-0.4123 &0.097 &  7.3 &  5.13 & 106\\
  605 &    18.8158 & 0.2135 &0.045 & 42.4 & 46.52 &  81\\
  625 &    27.0677 & 0.3292 &0.092 &600.2 & 519.95 &114\\
  656 &    43.3733 &-0.2349 &0.029 & $<$23.7 &  6.96 &1676\\
  \hline
\end{tabular}
\label{table1}

\medskip
\small{Column description: (1) name; (2)-(3) coordinates of the optical source;
  (4) redshift; (5)- (6) radio flux at 1.4 GHz from NVSS and FIRST
  surveys (mJy); (7) [O~III] emission line flux (10$^{-17}$ erg s$^{-1}$ cm$^{-2}$).}
\end{table}

%
%

\begin{table}
  \caption{Results of the JVLA observations}
\begin{tabular}{|l|c||c|c|c|c|}
  \hline
  ID  & Morph. & F$_{tot}$ & F$_{core}$ & beam  &  rms \\
  \hline
  {\bf 511}      &  SR & 1.15  &   &   1\farcs2 & 0.04    \\
                       & 1\arcsec &  0.34 &  0.29 &   0\farcs4 & 0.015 \\
                       & 1.8~kpc         &  0.26 &  0.26 &   0\farcs2 &  0.01 \\
\hline
 {\bf  519}     &  SR  & 12.21  &   9.4  &   1\farcs2 & 0.05   \\
                     &  0\farcs9     &  3.8   &   2.92 & 0\farcs4 & 0.04 \\
                      &  1.3~kpc    &  2.2   &   1.96 & 0\farcs2 &  0.013  \\
\hline
 {\bf  524}    & SR  &  25.42  &      &1\farcs2 & 0.04 \\
                     &   0\farcs3    & 13.21  &     &0\farcs4 & 0.05 \\
                     &  0.5~kpc      & 9.16  &      & 0\farcs2 &  0.05 \\
\hline
 {\bf 535 }     &  P &   4.91   &    & 1\farcs2 & 0.05   \\
                      &      &   2.24  &     &  0\farcs4 & 0.02  \\
                   &$<$0.3~kpc  &  1.86   &     &   0\farcs2 &  0.015  \\
\hline
 {\bf 537 }     &  P &  5.00  &         &  2\farcs2 x 1\farcs2 & 0.07  \\
                      &       &   3.57  &          &  0\farcs4 & 0.1 \\
                      & $<$0.2~kpc  &  2.99   &  2.99    &   0\farcs2  & 0.02 \\
\hline
 {\bf  547}     & \tiny{twosid. jet}   & 12.22  &        & 1\farcs7 x 1\farcs0  & 0.05  \\
                     & 2\arcsec &  4.30   &     & 0\farcs7 x 0\farcs5 & 0.02 \\
                     & 2.8~kpc          &  2.62    & 0.93  & 0\farcs2 & 0.01 \\
\hline
 {\bf  567}     &  P  &   3.30   &        &  2\farcs0 x 1\farcs0 & 0.04  \\
                      &     &   0.85   &          &   0\farcs4 & 0.02 \\
                     & $<$0.2~kpc  &   0.74   &          &   0\farcs2 & 0.01 \\
\hline
 {\bf  568}    & \tiny{double}  &  37.9  &       &  2\farcs6 x 2\farcs1 & 0.04  \\
                     &  1\arcsec & 27    &      & 0\farcs4 & 0.06  \\
                     &  0.8~kpc        &  13.2 &      & 0\farcs35 &  0.02 \\
\hline
 {\bf  590}  & \tiny{elongated}  & 5.9    &       &  1\farcs5 x 1\farcs1 & 0.03  \\  
                   &  0\farcs8 & 3.5   &       &  0\farcs4 & 0.03 \\
                   &  1.5~kpc       & 2.5   &       & 0\farcs25 &  0.02 \\  
\hline
 {\bf  605}  & P  & 44.3    &       &  2\farcs1 x 1\farcs3 & 0.02  \\  
                   &     & 42.1   &       & 0\farcs6 x 0\farcs5   & 0.02 \\
                   &  $<$0.9~kpc     & 36.5   &       & 0\farcs25 &  0.04 \\  
\hline
 {\bf  625}  & \tiny{hybrid}  & 420    &       &  1\farcs1 x 0\farcs9 & 0.03  \\  
                   & 22\arcsec    & 144   & 32.3       & 0\farcs6 x 0\farcs4   & 0.03 \\
                   &  37~kpc     & 103   &  31.5     & 0\farcs24 x 0\farcs21 &  0.02 \\  
\hline
 {\bf  656}  & SR  & 7.4    &       &  1\farcs8 x 1\farcs6 & 0.06  \\  
                   & 0\farcs8    &  --  & --    & -- & -- \\
                   &  0.5~kpc   & --  &   --    & -- & -- \\  
  \hline

\end{tabular}
\label{table2}

\medskip
\small{Column description: (1) name; (2) radio morphology (P stands
  for point-source and SR for slightly resolved), size of the source
  in arcoseconds and in kpc in the three following rows, (3) total
  radio flux (mJy) respectively at 1.4 GHz, 4.5 GHz and 7.5 GHz
  in the three following rows; (4) radio core flux (mJy)
  respectively at 1.4 GHz, 4.5 GHz and 7.5 GHz in the three following
  rows; (5) beam size (arcseconds) at 1.4 GHz, 4.5 GHz and 7.5
  GHz in the three following rows; (6) rms (mJy) at 1.4 GHz, 4.5 GHz
  and 7.5 GHz in the three following rows.}
\end{table}

\begin{table}
\caption{Radio flux ratio}
\begin{tabular}{l|c|c|c|c}
\hline
  ID  & F$_{FIRST}$/F$_{NVSS}$  & F$_{1.4 \,\ \rm GHz}$/F$_{NVSS}$  &  F$_{1.4 \,\ \rm GHz}$/F$_{FIRST}$ & $R$  \\
\hline
  511 & 0.49    & 0.36    & 0.73  & 0.08 \\
  519 & 0.89    & 0.66    & 0.74  & 0.11 \\
  524 & 1.05    & 1.02    & 0.98  & $<$0.37 \\
  535 & 0.65    & 0.68    & 1.04  & 0.26 \\
  537 & 1.00    & 0.77    & 0.77  & 0.46 \\
  547 & 0.89    & 1.14    & 1.28  & 0.09 \\
  567 & $-^a$   & $-^a$   & 1.15  & 0.26$^b$\\
  568 & 0.98    & 0.66    & 0.71  & $<$0.23 \\
  590 & 0.70    & 0.81    & 1.15  & 0.34 \\
  605 & 1.10    & 1.04    & 0.95  & 0.86 \\
  625 & 0.87    & 0.70    & 0.81  & 0.05 \\
  656 & $-^a$   & $-^a$ & 1.06  & $-$  \\
\hline
\end{tabular}
\label{table2bis}

\medskip
\small{Column description: (1) name; (2) flux ratio between FIRST and
  NVSS; (3) ratio between 1.4-GHz flux from JVLA observation and NVSS
  flux; (4) ratio between 1.4-GHz flux from JVLA observation and FIRST
  flux; (5) core dominance $R$, i.e. ratio between 7.5-GHz core flux
  and NVSS flux. $^a$ two point-like FIRST sources are blended into a
  single NVSS catalog entry and $^b$ the core dominance is estimated
  from the FIRST flux.}
\end{table}

\section{The JVLA observations}
\label{obs}

We obtained 6 hours of observations with the JVLA of NRAO\footnote{The
  National Radio Astronomy Observatory is a facility of the National
  Science Foundation operated under cooperative agreement by
  Associated Universities, Inc.}  with A array configuration in 3 days
on December 2012 and January 2013. We observed 12 objects in scans of
2 hours for each group composed of 4 sources. Each source was observed
for $\sim$20 minutes spaced out by the pointing to the phase
calibrators (J0022+0014, J0059+0006,J0217+0144, and J2337-0230). The
flux calibrator was 3C~48 observed for $\sim$6-7 minutes. The
observations were performed in L and C bands; the exposure time was
split in the two band configurations. While the L band configuration
corresponds to the default 1GHz-wide band centered at 1.4 GHz, the C
band was modified based on our purpose. We divide the available 2-GHz
bandwidth into two sub-band of 1 GHz centered at 4.5 and 7.5 GHz. This
strategy allows to obtain images in 3 different radio frequencies in
two integration scan. Each of the three bands was configured in 7
sub-bands of 64 channels of 1 MHz. The object ID~656 has been observed
only in L-band due to a telescope failure.

The data reduction was performed with {\it AIPS} (Astronomical Image
Processing System) package according to standard procedures.  The
final image was then produced from the calibrated data set by {\it
  CLEAN}ing to convergence. The final map was obtained using the task
{\it IMAGR} with a beam size ranging between $\sim$0\farcs2 and
$\sim$2\farcs6, according to the band, and different weights to
increase the signal to noise ratio. We self-calibrated the maps
  of the sources with the gain flux $\gtrsim$10 mJy. The core component
parameters have been measured with the task {\it JMFIT}.

The noise level was measured in background regions near the
target. The final noise is strongly affected (mainly in L band) by the
presence of interferences within the large bandwidth. At high
frequency some observations show a gain error in a few telescopes that
we were not able to recover in the calibration phase. Since in some
sources the total flux density is too low to properly self-calibrate
the data in gain ($\sim$10 mJy or less) and even in phase (a few mJy
or less), all these problems produced final images with a measured
noise level higher than the expected thermal noise.

In Tab.~\ref{table2} we collect the radio properties. In particular we give
the total and core (when visible in the radio images) flux
densities in the 3
observing bands.

\section{Results}
\label{results}

\subsection{Radio properties}

The FIRST radio maps of the selected objects show unresolved compact
radio sources on a scale of 5\arcsec, with the exception of ID~625
which is extended on a scale of $\sim$ 35\arcsec. 

In the JVLA observations, most of the sources appears unresolved or
slightly resolved down to a resolution of $\sim$0\farcs2 which
corresponds to 0.2--0.4 kpc. Some sources show instead radio
structures on a scale of 1-3 kpc: ID~547 (a twin-jet source), ID~568
(a double source and lacking of the radio core), and ID~590 (with an
elongated radio morphology on a scale of 0\farcs8, possibly due to a
bent two-sided jet structure). Finally, ID~625 shows a hybrid
radio-structure \citep{gopal00}, being a FR~I on the North side and a
FR~II on the South side on a scale of $\sim$40 kpc. Tab.~\ref{table2}
collects all the radio properties derived from the new
observations. Appendix~\ref{notes} provides additional comments on the
radio maps for the sources which show extended structures (Figures
from \ref{547} to \ref{625}). 

One of the main purposes of our program is to isolate the radio core
component and to measure the core dominance of the sources of our
sample. However, the high resolution JVLA observations required with
this aim and obtained with relatively short exposure time might be
missing faint extended emission. In order to explore this possibility
we compared the JVLA 1.4 GHz fluxes with those measured by FIRST and
NVSS (see Table \ref{table2bis}). The JVLA/FIRST flux ratios range
from 0.71 to 1.28, with an average value of 0.95 indicating that we
recovered most of the radio emission. The comparison with the NVSS
data is more complex due to the low resolution of this survey. Indeed,
in two cases (namely ID 567 and ID 656) two point-like FIRST sources
are blended into a single NVSS catalog entry. Leaving these two
sources aside, the JVLA/NVSS flux ratios cover a range between 0.36
and 1.14, with an average of 0.78. We conclude that our measurement
recover most of the total radio flux of our sources. In general, the
fraction of missing flux is sufficiently small and it does not to
alter our view of the overall radio structure.

\begin{figure*}
\includegraphics[scale=0.7]{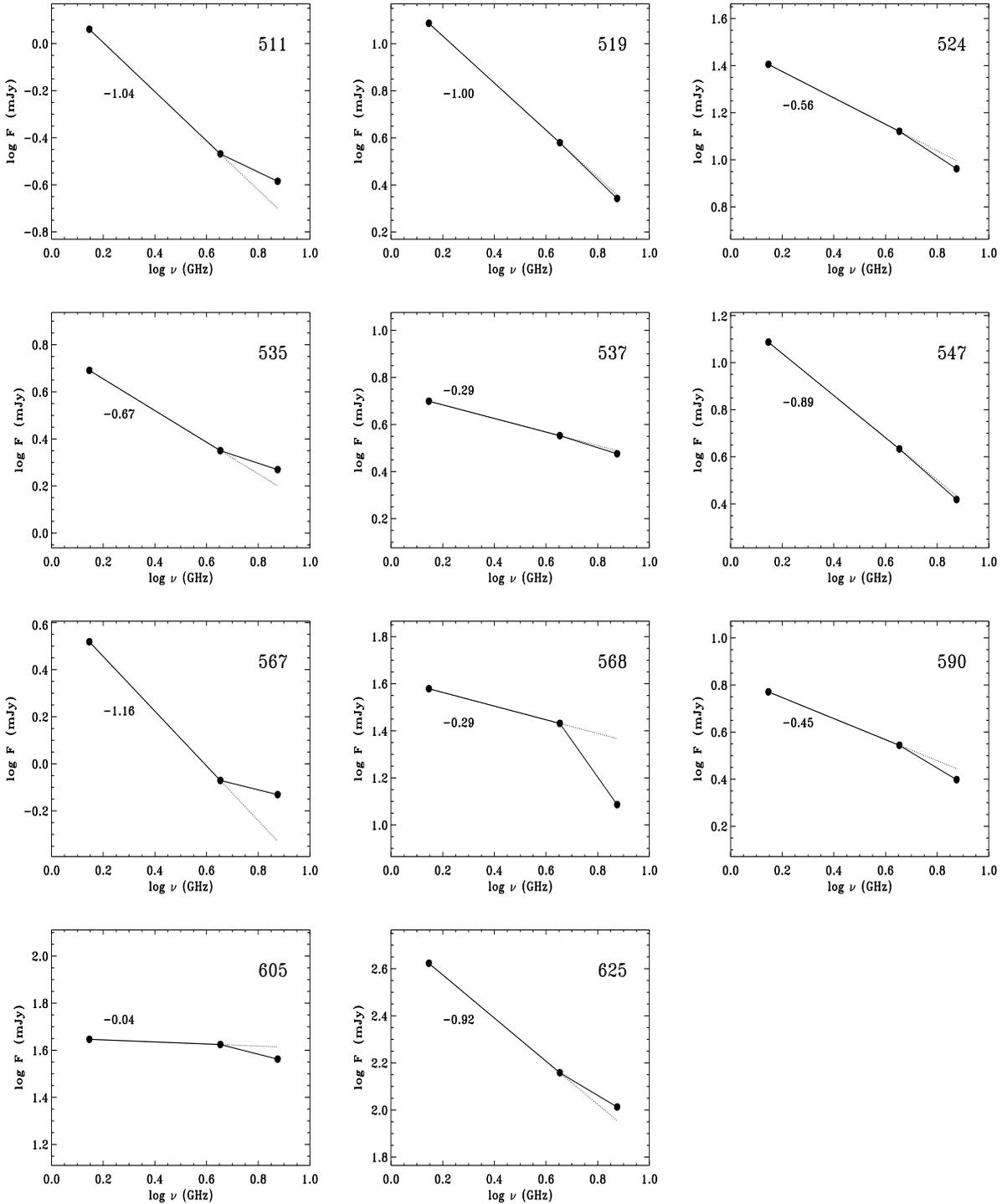}
\caption{Radio spectra of the 11 objects observed with JVLA observed at 1.4,
  4.5 and 7.5 GHz. The dotted line is the extrapolation of the 1.4-4.5 GHz
  slope which is reported in each panel.}
\label{SED}
\end{figure*}

We obtained matched-beam radio maps to derive the radio spectra at the three
frequencies (see Fig.~\ref{SED}). For the object ID~656 it is not possible to
derive its spectrum because of the lack of its C-band observation.

In Fig.~\ref{SED} we report the spectral slope $\alpha$ ($F_{\nu}
\propto \nu^{\alpha}$), obtained between 1.4 and 4.5 GHz.  The
spectral indices range from -1.16 to -0.04, indicative of both steep
and flat spectra. We note that source 568 shows a strong steepening at
7.5 GHz. This might be due to some missing flux at the highest
frequency, although we note that its double structures at 4.5 and 7.5
GHz are very similar. We suggest that the radio lobes spectra might be
intrinsically steep.

One of the key point of this study is the measurement of the radio
core. The high resolution radio maps at 7.5-GHz reveals a radio core
component for five sources (ID~511, 519, 537, 547, and 625). For the
remaining sources, we can use the radio spectra to measure the core
emission. Two sources (ID 590 and 605) show an overall flat ($\alpha
>$ -0.5) radio spectrum, consistent with a self-absorbed synchrotron
emission; the flux at 7.5 GHz is dominated by the radio core
emission. For ID~535 and 567 a spectral flattening occurs at 7.5 GHz
where the core emerges over the optically-thin radio emission. For the
last three objects of the sample we use the total flux at 7.5 GHz as
upper limit for the radio core measurement.

\begin{figure*}
\includegraphics[scale=0.7]{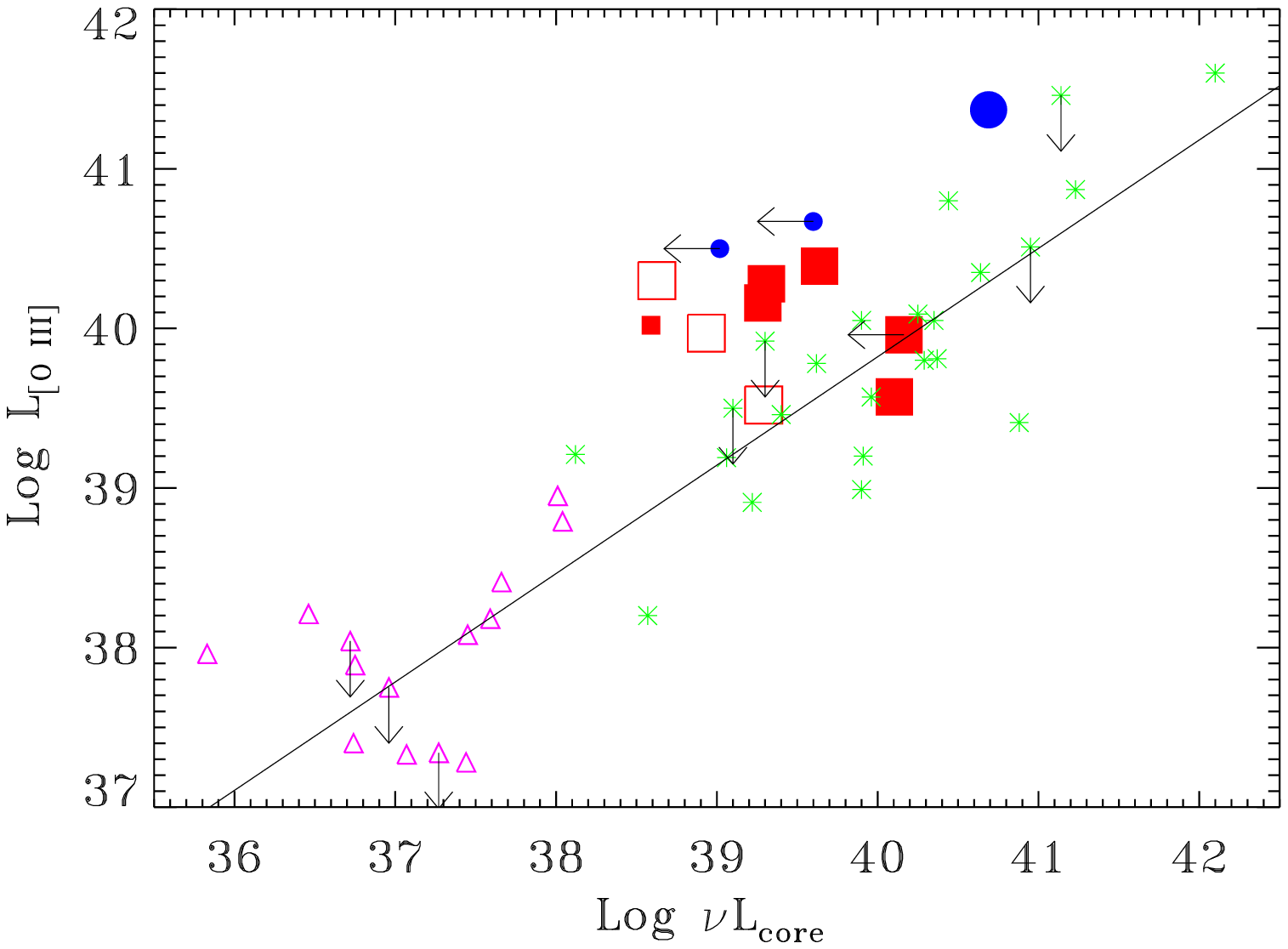}
\caption{Core radio power vs. [O~III] line luminosity (erg s$^{-1}$)
  for CoreG (pink triangles), 3CR/FR~I radio-galaxies (green stars),
  and our sample (colored symbols like in Fig.~\ref{diag}). The line
  indicates the best linear the for 3CR/FR~Is.}
\label{coreo3}
\end{figure*}

Having measured the radio core component thanks to the new observations, we
can include our sources in the $L_{core}$ vs. \loiii\ plane
(Fig.~\ref{coreo3}), similarly to what done for CoreG in \citet{baldi09}. The
radio core power distribution span a range similar to the local
3CR/FR~Is. Several sources lie in the same region of the 3CR/FR~Is also
considering their line luminosities, but others show a radio core deficit (or
line excess) with respect to the relation defined by the FR~Is. The sources
classified as LEGs, on average, lie closer to the relation than the HEGs.

The core dominance of our sources, $R$, is defined as the ratio
between the source nuclear emission at 7.5 GHz and the total flux
density, for which we adopt the NVSS measurement (Table
\ref{table2bis}). We preferred to use the 7.5 GHz core measurements,
with respect to the usual definition based on 5 GHz data, to limit as
much as possible the contamination from the extended structures. The
core dominance ranges between 0.05 and 0.86. The smallest R belongs to
the most extended object, the FR~I/FR~II ID~625.

\begin{table*}
\caption{Spectroscopic and photometric properties}
\begin{tabular}{l|c|c|c|c|c|c|c|c|c|c|c|}
\hline
  ID  &  opt. class  &  D$_{n}$(4000) &  C$_{r}$ &   M$_{r}$ & Log M$_{*}$  & Log M$_{BH}$  & Log L$_{\rm [O~III]}$ &  Log L$_{FIRST}$  & Log L$_{NVSS}$  & Log L$_{\rm core}$ & radio class \\
\hline
  511 & LEG  & 1.52  & 2.63 & -21.74 & 11.10  &    7.65 &    40.30    &  38.68   &   38.99   &    38.63    & RQ \\
  519 & LEG   & 1.74  & 2.42 & -22.11 & 11.30  &    8.67 &    40.28   &   39.50  &   39.56   &     39.31  & FR~0  \\
  524 & LEG  & 1.89  & 2.87 & -21.50 & 11.07  &    8.32 &    39.96   &   39.89   &   39.87   &     $<$40.16& FR~0  \\
  535 & LEG  & 1.75  & 2.83 & -22.61 & 11.48  &    8.76 &    40.26   &   38.96   &   39.14   &     39.29   & FR~0  \\
  537 & LEG  & 1.95  & 2.87 & -21.38 & 11.09  &    7.72 &    39.52   &   38.90   &   38.90   &     39.29   & FR~0    \\
  547 & LEG  & 1.67  & 3.57 & -21.62 & 11.18  &    7.64 &    39.97   &   39.22   &   39.27   &     38.93   & FR~0     \\
  567 & LEG  & 1.44  & 2.69 & -20.95 & 10.62  &    7.14 &    40.02   &   38.45   &   $<$38.91   &     38.59   & RQ    \\
  568 & HEG  & 1.56  & 2.33 & -20.99 & 10.76  &    7.19 &    40.67   &   39.50   &   39.51   &     $<$39.60& RQ      \\
  590 & LEG   & 1.93  & 3.29 & -22.47 & 11.42  &    8.43 &    40.39   &   39.22  &   39.37   &     39.64  & FR~0      \\
  605 & LEG  & 1.99  & 3.10 & -21.86 & 11.15  &    8.57 &    39.57   &   39.47   &   39.44   &     40.10   & FR~0      \\
  625 & HEG  & 1.61  & 3.17 & -22.13 & 11.24  &    8.18 &    41.37   &   41.18   &   41.24   &     40.69   & FR~I/FR~II     \\
  656 & HEG  & 1.39  & 3.16 & -19.25 &  9.97  &    7.08 &    40.50   &   38.26   &   $<$38.79   &     $<$39.02&  RQ      \\
\hline
\end{tabular}
\label{table3}

\medskip
\small{Column description: (1) name; (2) optical spectroscopic
  classification; (3) color based on D$_{n}$(4000) (typical error
    0.07$-$0.14); (4) concentration index (typical error
    0.01$-$0.03); (5) M$_{r}$ absolute r-band magnitude 
    (typical error $<$0.02); (6) galaxy stellar mass M$_{*}$
  (M$_{\odot}$) (typical error 0.15); (7) black hole mass
  M$_{BH}$ (M$_{\odot}$) (typical error 0.07$-$0.12); (8)
  [O~III] luminosity (erg s$^{-1}$) (typical error 3$-$10\%);
  (9) 1.4 GHz FIRST luminosity (erg s$^{-1}$); (10) 1.4 GHz NVSS
  luminosity (erg s$^{-1}$) used as total radio luminosity L$_{\rm
    tot}$; (11) 7.5 GHz VLA luminosity (erg s$^{-1}$) used as radio
  core luminosity L$_{\rm core}$; (12) radio class: radio-quiet (RQ)
  AGN, FR~0, hybrid FR~I/FR~II.}
\end{table*}

\subsection{Spectro-photometric properties}

We now explore the physical properties of the present objects 
based on the optical spectroscopic
and photometric information available from the SDSS survey. We consider the
mass of their central black hole (BH), the host mass and morphology, and the
$D_n(4000)$ index (see Tab.~\ref{table3}), similarly to the study of SDSS/NVSS
sample by \citet{baldi10a}.

We estimate the BH mass ($M_{\rm{BH}}$) from the stellar velocity dispersion
adopting the relation of \citet{tremaine02}. They range from $\sim$10$^{7}$ to
$\sim$10$^{9}$ $M_{\odot}$. The sources are associated with galaxies with a
distribution of masses\footnote{$M_{*}$ is estimated from the $g$ and $z$ SDSS
  magnitudes following following \citet{bell07} resulting in slightly
  different masses from those reported by \citet{best05a}.} in the range
$10^{10}$--$10^{11.5}$ M$_{\odot}$ (and a median of $\sim 10^{11}$ M$_{\odot}$).

\begin{figure*}
\centerline{
\includegraphics[angle=90,scale=0.4,angle=0]{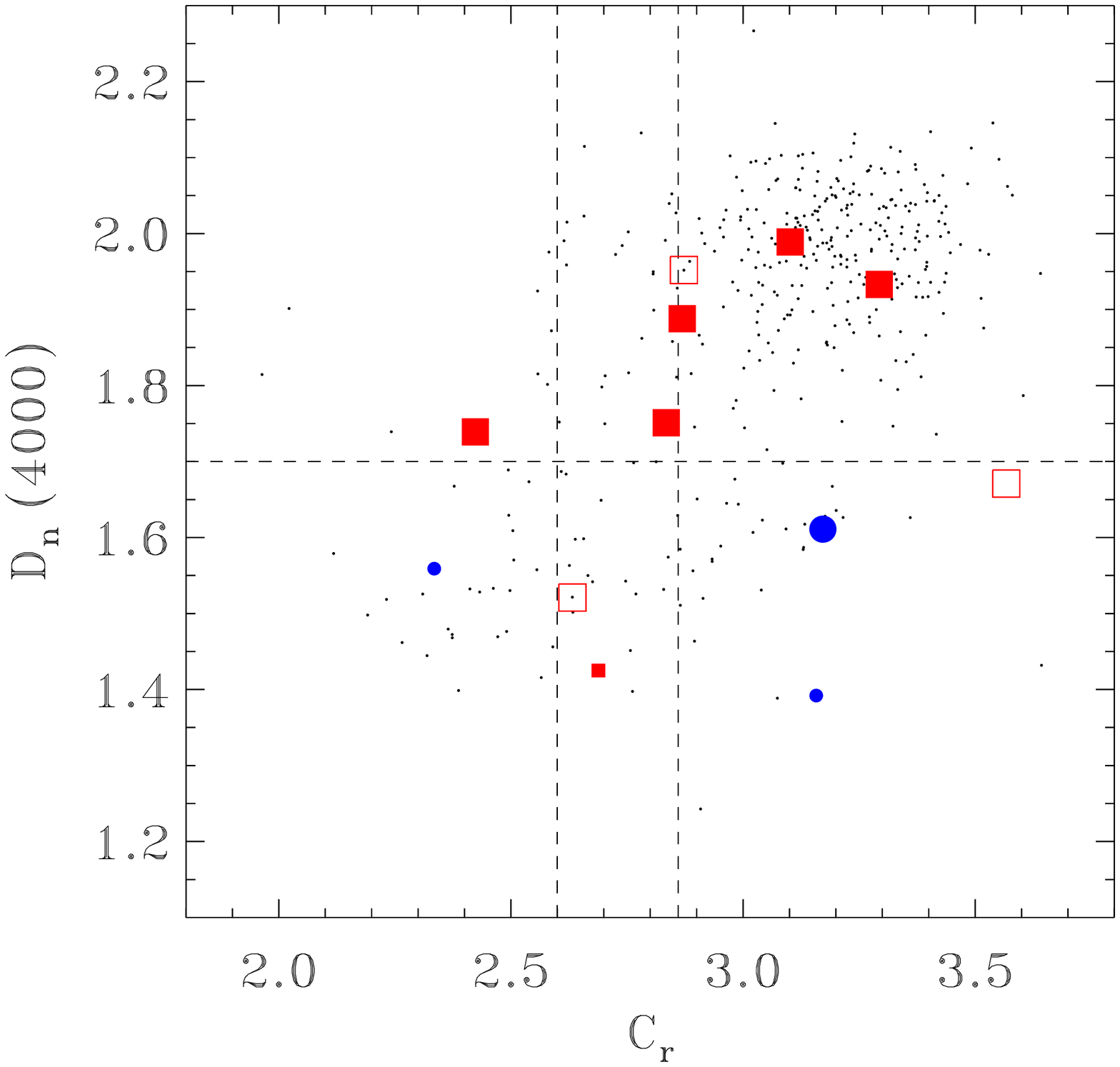} 
\includegraphics[angle=90,scale=0.4,angle=0]{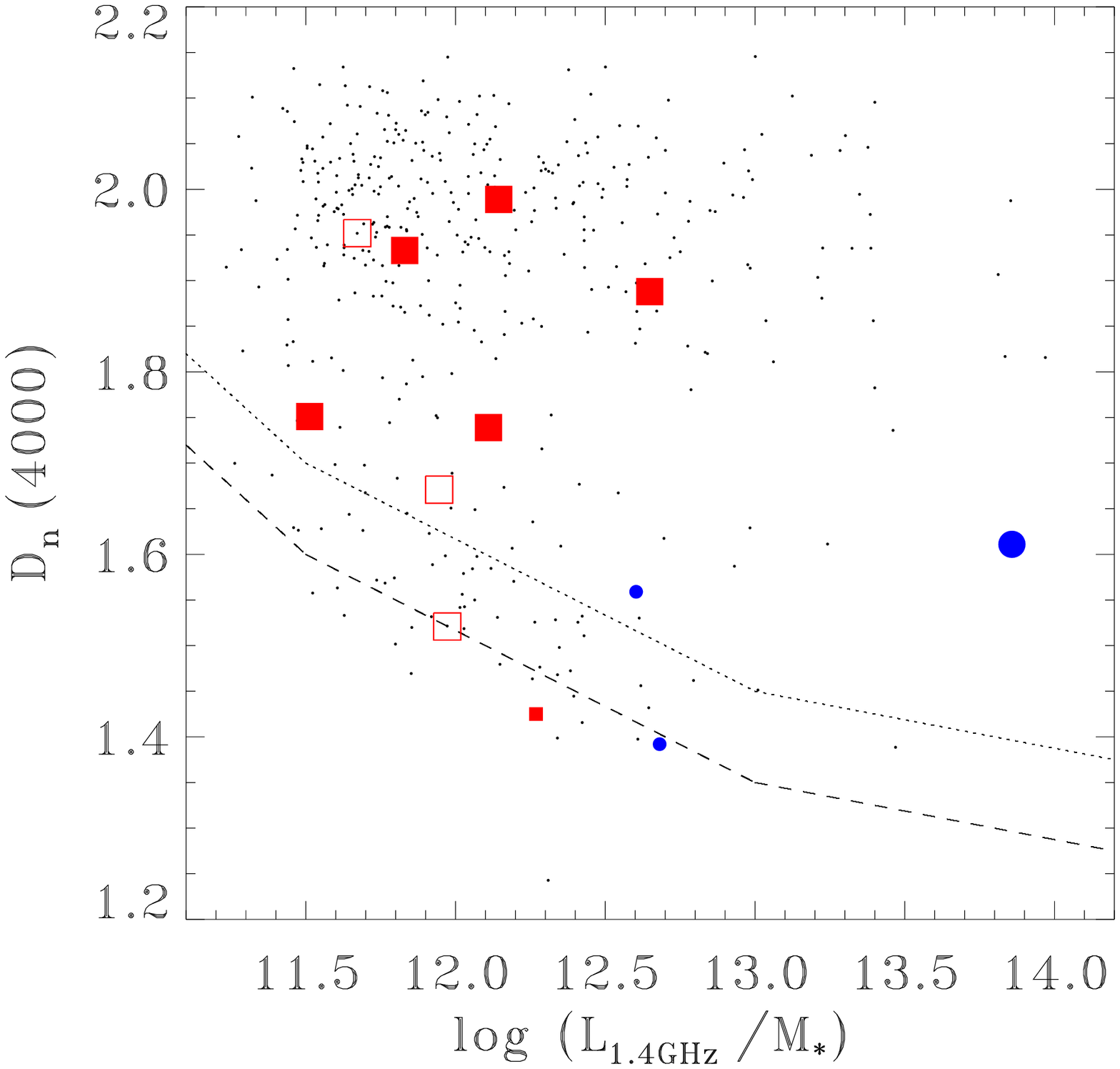}}
\caption{Left panel: concentration index in the r band, $C_r$, versus the
  $D_n(4000)$ index. The vertical lines correspond to the two
  definition of ETGs ($C_{r}>$ 2.86 or 2.60, see text for details), while the
  horizontal line separates the blue from red sources. Right panel:
  $D_{n}(4000)$ versus $L_{\rm 1.4 \,\,GHz}/M_{*}$. The dashed
  curve is the empirical separation between the star forming and AGN radio
  emitting sources, performed by \citet{best05a} (read the text for
  details). The dotted curve is shifted 0.1 above in $D_{n}(4000)$ of the
  dashed curve, in order to separate the objects with a possible 
  star-formation contribution to their radio emission.  The small dot points
  correspond to the SDSS/NVSS sample studied by \citet{baldi10a}. The
  different color/shape symbols correspond to the different classes like in
  Fig.~\ref{diag}.}
\label{spdd}
\end{figure*}

We then study the host type by using the concentration index $C_{r}$ (defined
as the ratio of the radii including 90\% and 50\% of the light in the $r$ band
respectively). ETGs have $C_{r}\geq 2.86$
(e.g. \citealt{strateva01,kauffmann03b,bell03}). This cut basically
corresponds to E type with a small contamination of S0 and Sa
\citep{bernardi10}. A less conservative selection of ETGs uses $C_{r}\geq
2.60$ (e.g. \citealt{nakamura03,shen03}), which yields a larger contribution
from S0 and Sa.

The 4000\AA\ break strength, $D_n(4000)$ (defined as the ratio between
the fluxes in the wavelength range 3850-3950\AA\ and 4000--4100\AA) is
instead sensitive to the presence of a young stellar population
\citep{balogh99}.  Massive galaxies (log $M_*/M_{\odot} \gtrsim 10.5$)
shows a bimodal distribution of $D_n(4000)$ \citep{kauffmann03b}. The
transition between the two groups is located approximately at
$D_n(4000) \sim 1.7$ with the red (blue) hosts having larger (smaller)
values.

A further diagnostic panel tool enables us to to qualitatively measure the
amount of contamination in radio emission in the galaxy due to star
formation. This method is based on the location of a galaxy in the
$D_{n}(4000)$ versus $L_{{\rm 1.4 \,\,GHz}}/M_{*}$ plane, where $M_{*}$ is the
galaxy's stellar mass.  \citet{best05a} predict the radio emission per
unit of galaxy mass expected from a stellar population of a given
age: when a galaxy shows an excess larger than 0.225 in
$D_{n}(4000)$ above the curve corresponding to the prediction of a star
formation event lasting 3 Gyr and exponentially decaying, its radio emission
is associated with the AGN.

Figure~\ref{spdd}, in the left panel, shows the distribution of the objects in
the spectro-photometric diagram composed of $C_{r}$ vs. $D_n(4000)$,
while in the right panel we compare $L_{\rm 1.4 \,\,GHz}/M_{*}$ and
$D_{n}(4000)$.  In order to explore the properties of the sample, we
separated the sources into three groups based on the BH masses: large
$>$10$^{8}$ M$_{\odot}$, intermediate $\sim$10$^{7.7}$ M$_{\odot}$ and
small $<$10$^{7.3}$ M$_{\odot}$. 

We note a different behavior of the sources in our sample, related with their
BH mass. A first group consists in 4 sources with small (and one intermediate)
BH masses, namely ID~511, 567, 568, and 656. Three of them (see
Fig. ~\ref{spdd}, right panel), have a likely contribution of a star formation
to their radio emission; this is supported also by their large NVSS/FIRST
radio flux ratios, an indication of the presence extended diffuse radio
emission, missed by the FIRST images. The fourth object (ID~568) is consistent
with being a Seyfert 2 galaxy, based on its location in Fig.~\ref{lrlo3} and
on its spectrophotometric properties.  They are all blue galaxies, having
$D_{n}(4000) < 1.6$.  In the diagram comparing the core power, L$_{\rm core}$,
and emission line luminosity (Fig.~\ref{coreo3}) they all show a large deficit
of radio core emission, typically a factor of $\gtrsim$100 , with respect to the
relation defined by 3CR/FR~Is.

Conversely, the second group includes sources with large/intermediate
BH masses, whose radio emission is dominated by the AGN. With the sole
exception of ID~625 (the hybrid FR~I/FRII radio source) they are
associated with red massive ETGs belonging to the LEG spectroscopic
class. With respect to the first group, they follow more closely the
L$_{\rm core}$ vs \loiii\ relation. 

\begin{figure}
\includegraphics[scale=0.45]{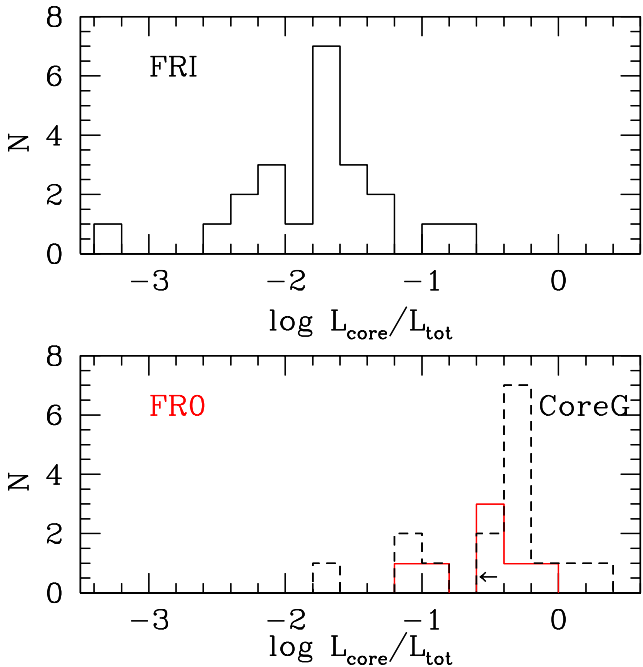}
\caption{Histograms of core dominance Log R = log (L$_{\rm
    core}$/L$_{\rm tot}$) for the 3CR/FR~Is (black solid line), CoreG
  (black dashed line) and our sample of seven FR~0s (red solid line).}
\label{cd}
\end{figure}

Fig.~\ref{cd} shows that the core dominance for this sub-sample ranges
from $\log R \sim$ -1 to $\sim 0$ with a mean value of $\sim$-0.5.
This distribution is significantly different from from that of
3CR/FRIs (with a $>$99.9\% probability, according to a Kolmogorov
Smirnoff test) while it is not distinguishable from the $R$
distribution of CoreG.  We remind that we estimated $R$ as the ratio
between the 7.5-GHz core emission and the total 1.4-GHz flux (NVSS),
while for the 3CR/FRIs and CoreG we used the 5 GHz core flux against
the total 1.4-GHz flux (NVSS).  However, since the radio core emission
has generally a flat spectrum this quantity is only weakly dependent
on the frequency used for the core measurement and this comparison is
robust.

\section{Discussion}
\label{discussion}

The results discussed in the previous sections indicate that the
twelve objects, belonging to the SDSS/NVSS sample and observed with
the JVLA, can be divided in two main groups, corresponding to specific
radio, spectroscopic, and photometric properties. Before discussing
these groups in more detail, we treat here the only object that shows
rather different properties. Not surprisingly, by exploring a sample
of radio-sources, we found a classical extended RG, namely ID~625. As
already mentioned it has a hybrid FR~I/FRII morphology extending over
$\sim$40 kpc. All its characteristics are typical of FR~II sources
considering its BH mass, the blue host color, its spectroscopic class
(HEG), and its location in both the L$_{\rm core}$ and L$_{\rm tot}$
vs \loiii\ diagram in the region populated by the FR~II part of 3C
sample \citep{chiaberge02,buttiglione10,chiaberge11}.

The first main group consists of four sources which are characterized
by their low BH masses, mostly $\sim$10$^{7}$ M$_{\odot}$ and their
blue color. Their radio and spectrophotometric properties indicate
they are RQ AGN (Tab~\ref{table3}). The presence of an active nucleus
is witnessed by their optical line ratios and equivalent widths,
characteristic of AGN. Nonetheless, three of them shows a substantial
contamination from star formation to their radio emission.  The fourth
object (ID~568) shows a double radio morphology with a size of 0.8 kpc
and a steep spectrum, often observed in Seyfert galaxies
\citep{edelson87,nagar05}. In addition, its radio and line power
locate this source in the area typical of the Seyfert galaxies. The
contamination from RQ AGN ($\sim$33\% of the sample) is higher than
observed in the whole SDSS/NVSS sample \citep{baldi10a}. This is due
to the fact that the selection criterion of this pilot sample based on
the high [O~III] equivalent width favors the inclusion of bright
emission-line galaxies, which are preferentially RQ AGN.

The second group consists of seven sources, representing (together
with the powerful source ID~625 discussed above) the genuine RL AGN
population in the SDSS/NVSS sample. They live in red massive
($\sim$10$^{11}$ M$_{\odot}$) ETGs, have BH masses $\gtrsim$10$^{8}$
M$_{\odot}$ and are spectroscopically classified as LEGs (apart from
the HEG-spectrum FR~I/FR~II 625). All these properties are shared with
FR~I RGs (e.g.,
\citealt{zirbel96,cao04,floyd08,chiaberge99,baldi08,baldi09,buttiglione10}). Furthermore,
their radio core and [O~III] luminosities lie in the range typical of
FR~Is. Therefore, all the nuclear and host characteristics of this
group are indistinguishable from those of FR~Is.  Nonetheless, the
JVLA observations presented here show compact radio structures (with a
limit to their size of $\sim$ 0.5 kpc or, in a few cases, extending by
at most 1-3 kpc) and lead to an estimate of their (average) core
dominance a factor of $\sim$30 higher than FR~Is.  The only feature
distinguishing them from FR~Is is then the substantial lack of
extended radio emission. For such a paucity we define them ’{\it
  FR~0}’.

\subsection{The FR~0 population}
\label{FR0}

The FR~0 classification corresponds to a combination of radio and
spectro-photometric properties. In the radio band, FR~0s are
characterized by a core responsible for a fraction larger than
$\gtrsim$10\% of the total emission with structures extending for, at
most, a few kpc.  In classical FR~Is and FR~IIs RGs the core dominance
is typically $\lesssim$1\% and their sizes are of tens or hundreds of
kpc. (e.g, \citealt{morganti93,morganti97}). The high FR~0 core
dominance appears to be due to the paucity of extended radio emission,
rather than to an enhanced radio core, since FR~0s and FR~I show
similar ratios of radio-core to emission line luminosity. In addition,
the FR~0 definition involves the photometric/spectroscopic nuclear and
host properties (e.g., BH mass, color, and emission line ratios,
similar to FR~Is), which distinguish them from the radio-sources
associated with RQ AGN and star forming galaxies.

This new class of RGs is similar to the low-luminosity radio sources
hosted in ETGs, studied by \citet{slee94}, which contain parsec-scale
radio cores and do not produce extended radio emission. Furthermore,
recently, \citet{sadler14} found that the bulk of the 20-GHz RG
population consists of compact radio sources lacking of extended radio
emission, analogous to our FR~0s.

The CoreG fulfill the requirements for a FR~0 classification: they show kpc
scale radio structures and are of high core dominance, they are hosted in red
giant ellipticals and are characterized by LEG line ratios.  They are
$\sim$100 times less luminous than the FR~0s of the SDSS/NVSS sample discussed
here. They smoothly extend the various nuclear multi-wavelengths relations
seen in FR~Is. In this sense
they represent the low luminosity end of the FR~0 population that therefore
extends at least down to a radio power of $\sim 10^{36}$ $\ergs$.

The small size and the steep spectra of FR~0s might suggest the
presence of Compact Steep Sources (CSS), i.e. of rapidly growing young
radio-sources (e.g. \citealt{odea98,snellen00}). CSS show a
correlation between the turnover frequency in their spectra and their
linear sizes (e.g., \citealt{fanti90,odea98}). The sizes of our FR~0
sources ($<$3 kpc) does not contrast with their steep spectra since
the turnover might occur at lower frequencies than the lowest value
sampled by our observations, 1.4 GHz. However, the core dominance of
CSS, although generally more difficult to measure and uncertain with
respect to more extended radio-sources, is much lower than in FR~0s
(typically ${\rm log} \,R \sim$-1 for quasars and $\log \,R \sim$-2.4
for galaxies \citealt{saikia95,saikia01}) ruling out a general
association of the FR~0s with the CSS.

As reported in the Introduction, the vast majority of the SDSS/NVSS
objects fulfill the FR~0 definition with the notable exception of the
lack of a direct measurement of the core dominance, impossible from
the FIRST images. Indeed, this was the main motivation to obtain the
higher resolution data presented in the previous
sections. Nonetheless, the possibility that they are generally high
core dominance objects, and thence bona-fide FR~0, appears to be
reasonably founded. In fact, most of them ($\sim$80 \%) are unresolved
or barely resolved in the FIRST maps, limiting their sizes to, at
most, a few kpc. This is an indication that, similarly to the 7
objects studied, they have a deficit of extended radio emission with
respect to classical extended RGs. It is clearly highly desirable to
obtain proper radio data for a larger sample to base this conclusion
on firmer ground. Therefore, if this result was confirmed for a larger
sub-sample of the SDSS/NSS RGs, the FR~0 population would turn out to
be the dominant class of RGs in the local Universe with a space
density $>$100 times higher than 3C sources, as also suggested by
\citet{sadler14}.

FR~0s apparently differ from FR~Is just for the lack of large scale radio
structures. In the following we present two scenarios that can account for
this specific property.

In the first scenario, the central engines of FR~0s and FR~Is are
indistinguishable and the paucity (and small size) of the extended
radio emission is ascribed to an evolutionary effect. FR~0s might be
young RGs that did not yet developed an extended radio
structure. However, their number density should be smaller than that
of FR~Is, the opposite of what is observed. This scenario can hold if
FR~0s are intermittent sources. Rapid AGN intermittency (see, e.g.,
\citealt{readhead94,reynolds97,czerny09}) would prevent FR~0s from
becoming well developed RGs. However, this does not explain why
intermittency should affect only FR~0s, particularly considering the
indications that accretion in FR~Is and CoreG is associated with a
long-lasting inflow of hot coronal gas \citep{allen06,balmaverde08},
related only to the host properties. Therefore, no differences would be
expected between FR~0s and FR~Is.

As a second scenario, the differences between FR~0s and FR~Is might be
due to a lower jet bulk speed $\Gamma$ in FR~0s than in FR~Is. In this
scheme, the innermost regions of FR~0 and FR~Is do not differ
significantly since they share a common range of accretion rates
(proved by the similarity in line emission luminosity and optical line
ratios) and of radio core powers (which represent the synchrotron
emission from the base of the jet). The different radio behavior
should arise on a larger scale. In the hypothesis that jets in FR~0s
are slower than FR~Is, they are more subject to instabilities and
entrainment \citep{bodo13} and this causes their premature
disruption. Indeed, the typical scale of the radio emission in FR~0s
is smaller than the core size of their hosts, a region characterized
by a dense interstellar medium that obstructs the passage of the
jet. This idea is supported, albeit with a small number statistics, by
the absence of one-sided kpc scale morphologies among the FR~0s
observed, the typical sign of relativistic jet boosting.  The ultimate
origin of the lower $\Gamma$ factor in FR~0 is apparently not related
to any directly observable quantity. We speculate that this could be
due to a different spin of their central BH, by assuming a dependence
between the BH spin and $\Gamma$, as suggested by, e.g.,
\citet{mckinney05}, \citet{tchekhovskoy10}, \citet{chai12}, and
\citet{maraschi12}. The FR~I radio morphology is produced only when
the BH spin is close to its maximum value, while smaller spin values
could be associated with FR~0s.  Recently, \citet{ghisellini14}
predicted a population of jetted sources which have smaller BH spin
than classical RL AGN: the FR~0s might represent or be part of this
population.

\section{Summary and Conclusions}
\label{conclusion}

We present the observations of 12 objects with the JVLA in A-array
configuration. These sources are selected from a large sample
(cross-matching NVSS/FIRST and SDSS) which represents the bulk of the
local radio-emitting AGN population \citep{baldi10a}. Based on the
FIRST radio maps, most of the sources are compact on an angular scale
of 5\arcsec. 

The new high-resolution observations at 1.4, 4.5, and 7.5 GHz reveal
that they still show a compact morphology, extending at most over 1-3
kpc (with just one exception, a hybrid FR~I/FR~II which extends over
$\sim$40 kpc). We isolate and measure the radio core component thanks
to the new high-resolution maps or to the radio spectra. Furthermore,
the SDSS survey provides with the spectro-photometric properties of
the sample, such as, BH mass, host type, emission lines. Based on
these properties, we divide the sample into two groups.

The first group consists of four sources which represent the RQ AGN
contamination to the SDSS/NVSS sample. They have small BH masses (mostly
$\sim$10$^{7}$ M$_{\odot}$) associated with blue galaxies which show evidences
of star formation. 

The second group consists of seven RL AGN in red massive ($M_{*}
\sim$10$^{11}$ M$_{\odot}$ and BH masses $\gtrsim$10$^{8}$
M$_{\odot}$) ETGs spectroscopically classified as LEG. All these
characteristics are shared by local FR~Is.  In particular, the members
of this group have similar radio and [O~III] line luminosities of
FR~Is, lying on the same relation; however, they show a core dominance
higher than in FR~Is by a factor $\sim$30. What distinguishes this
group from FR~Is is one single aspect: the lack of substantial
extended emission. For this characteristic we define these objects
`FR~0'. This new RG class is consistent with a sub-class of
Gigahertz-Peaked Spectrum (GPS) radio sources, proposed by
\citet{tingay15}, with low luminosity and with jet-dominated compact
morphologies similar to FR~Is and lacking of extended radio
emission. Furthermore, recently, \citet{sadler14}, studying the
  20-GHz radio-source population in the local Universe, suggest that
  the FR~0s might be a mixed population of young CSS and GPS radio
  galaxies.

What causes the deficit of extended radio emission in FR~0s?  We
discuss two possible interpretations. The first scenario proposes that
the FR~0s consist in young intermittent radio sources that will
eventually evolve into extended RGs. However, this contrasts with the
picture in which low luminosity RGs are powered by continuous
accretion of hot gas from their X-ray coronae. Furthermore, it does
not explain why the intermittency should only occur in FR~0s and not
in FR~Is, considering the similarity of their nuclear and host
properties.

The second scenario suggests that FR~0s have smaller jet Lorentz factors
$\Gamma$ than FR~Is: their jets are less stable against entrainment and their
passage through the dense ISM region of the central regions of their hosts
causes their premature disruption. This low-$\Gamma$ scenario can be tested by
looking at the asymmetry between the opposite jets in FR~0s, a parameter
directly linked to the jet speed. Although twin symmetric jets are common in
FR~Is, they often have one jet substantially brighter than the other on kpc
scale (e.g., \citealt{parma87}). This indicates that the high Doppler factors
derived for FR~Is on pc scale (based on the detection of superluminal motions,
e.g., \citealt{giovannini01}, and, indirectly, on the association with BL~Lac
objects in the AGN unified model, e.g. \citealt{urry95}) are preserved to
larger scales. In the present sample we find that all three FR~0s with an
extended radio structure are two-sided and rather symmetric. The number is too
small to draw any firm conclusion (and two-sided jets are expected to be more
common than a one-sided morphology in a randomly oriented sample) but it is
consistent with this hypothesis.  We speculate that the ultimate difference
between FR~0s and FR~Is is the spin of their BH, being smaller for FR~0s.

High resolution imaging of a larger sample of local low luminosity
radio sources is needed to put this result on firmer statistical
ground and confirm the presence of the FR~0 population. In order to
improve the efficiency of such a survey, the results obtained in this
study indicate that it should be focused on objects with a LEG
spectrum, hosted in red early type galaxies. If the FR~0 population
was confirmed, they would represent the bulk of the RL AGN in the
local Universe with a large impact on our view of the low-luminosity
radio-emitting population.

\appendix
\section{Notes on the extended radio-sources}
\label{notes}
We report here notes on the structure of the extended radio-sources in the
sample.

\begin{figure*}
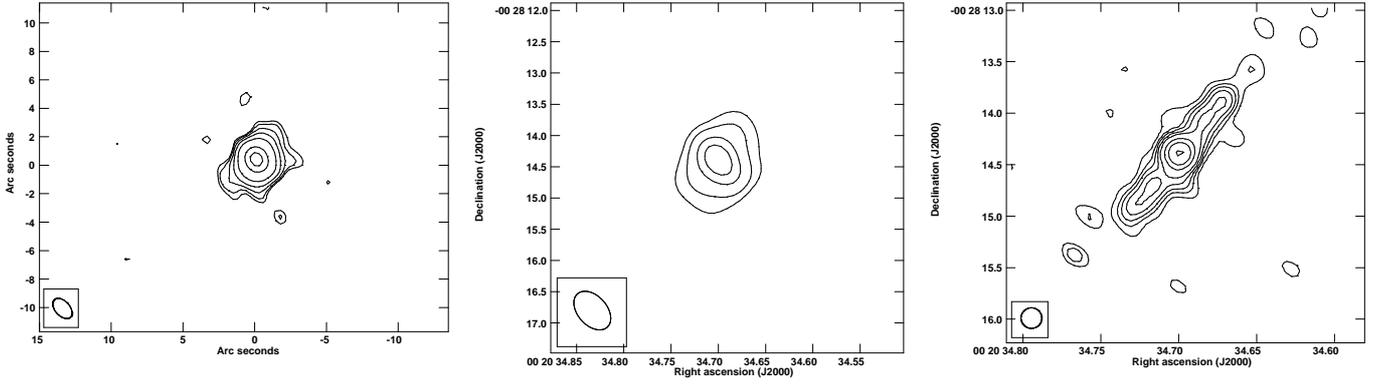

\centerline{
\includegraphics[scale=0.30,angle=270]{i547-l.ps}
\includegraphics[scale=0.30,angle=270]{i547-4gh.ps}
\includegraphics[scale=0.30,angle=270]{i547-7gh.ps}
}
\caption{JVLA images of ID 547 at 1.4, 4.5 and 7.5 GHz. The HPBWs and noise
levels are reported in Tab.~\ref{table2}. Levels are respectively: 0.15 0.2 0.3 0.5 1 3 5 mJy/beam; 0.3 0.5 0.9 1.3 mJy/beam; 0.03 0.05 0.07 0.1 0.15 0.3 0.5 0.9 mJy/beam}
\label{547}
\end{figure*}

{\bf ID 547}: while at 1.4 and 4.5 GHz it is only marginally resolved, a
clear twin-jet morphology is revealed at the higher frequency
(Fig.~\ref{547}). The two jets are rather straight and symmetric,
extending for for about 1\arcsec\ on each side of the nucleus along PA
$\sim$ -40$^\circ$.

\begin{figure*}
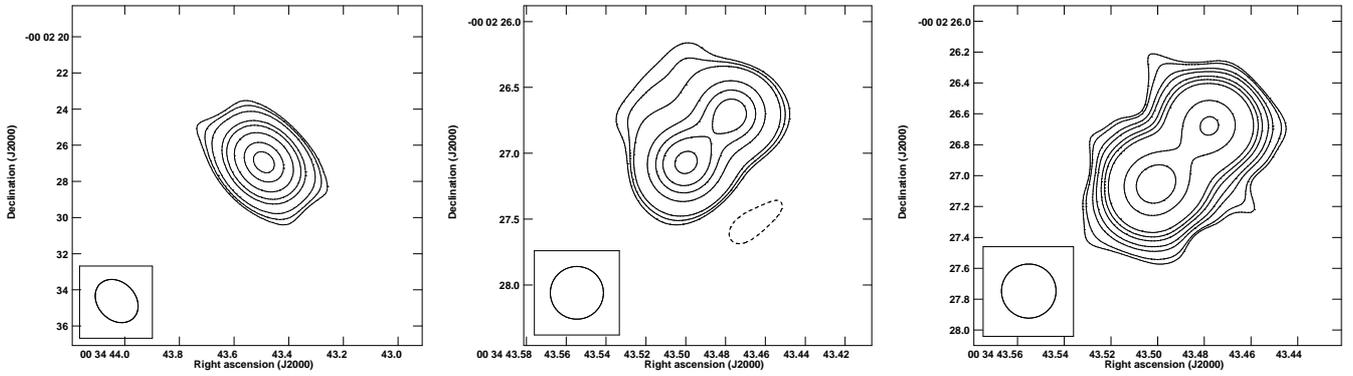

\centerline{
\includegraphics[scale=0.30,angle=270]{568-l.ps}
\includegraphics[scale=0.30,angle=270]{568-4.ps}
\includegraphics[scale=0.30,angle=270]{568-7.ps}
}
\caption{JVLA images of ID 568 at 1.4, 4.5 and 7.5 GHz. The HPBWs and noise
levels are reported in Tab.~\ref{table2}. Levels are respectively: 
0.3 0.5 1 3 5 10 20 30 mJy/beam; -0.5 0.5 0.7 1 3 5 7 10 20 30 50 70 mJy/beam;
0.07 0.1 0.2 0.3 0.5 0.7 1 3 5 mJy/beam}
\label{568}
\end{figure*}

\noindent
{\bf ID 568}: its radio morphology is double ($\sim$0\farcs5 in size) with PA
= $\sim$130$^{\circ}$ (Fig.~\ref{568}). The position of the two lobes are
(00$^{\rm h}$ 34$^\prime$ 43$\farcs$51 -00$^{\rm d}$ 02$^\prime$ 27$\farcs$06)
and (00$^{\rm h}$ 34$^\prime$ 43$\farcs$48 -00$^{\rm d}$ 02$^\prime$ 26$\farcs$67) whose
fluxes at 4.5 and 7.5 are respectively: 15.3 and 6.7 mJy for the former
component and 8.4 and 5.3 mJy for the latter. The resolution is insufficient
to resolve sub-structures or the presence of a radio core.

\begin{figure*}
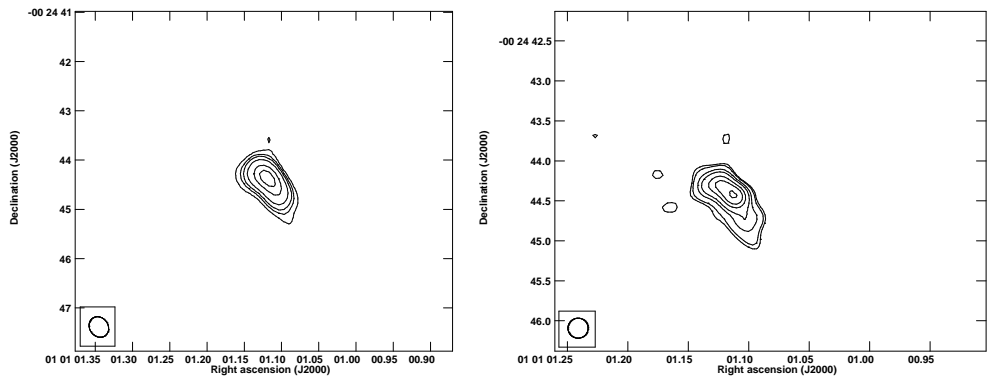

\centerline{
\includegraphics[scale=0.30,angle=270]{590-4.ps}
\includegraphics[scale=0.30,angle=270]{590-7.ps}
}
\caption{JVLA images of ID 590 at 4.5 and 7.5 GHz. The HPBWs and noise
levels are reported in Tab.~\ref{table2}. Levels are respectively: 
0.1 0.2 0.3 0.5 1 1.5 mJy/beam; 0.07 0.1 0.2 0.3 0.5 0.7 1 mJy/beam}
\label{590}
\end{figure*}

\noindent
{\bf ID 590}: the source is point-like in L band, while at high frequencies
the source is elongated with PA=45$^{\circ}$ with a size of 0\farcs8
(Fig.~\ref{590}) with a morphology suggestive of a bent two-sided jet
structure.

\begin{figure*}
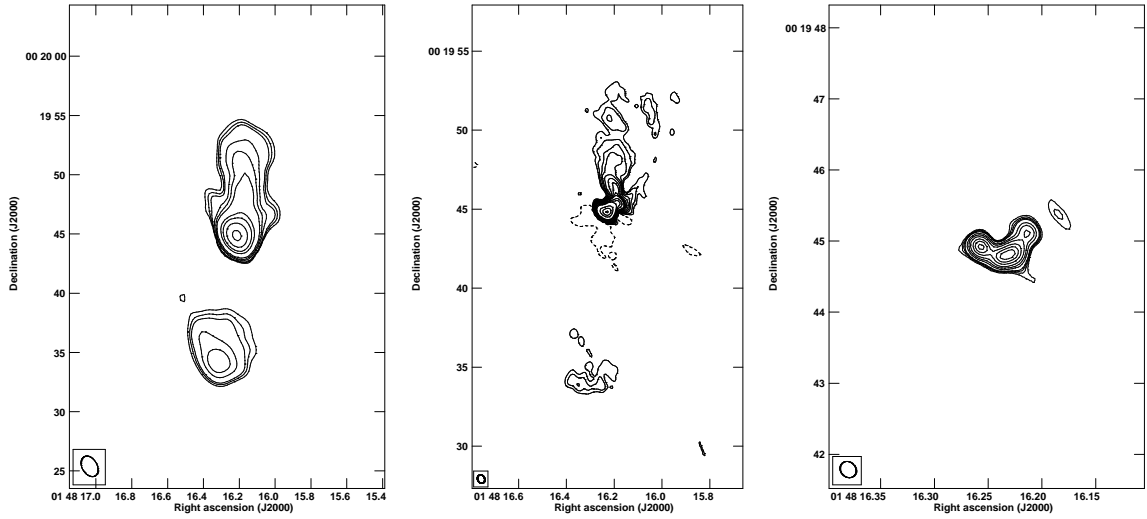

\centerline{
\includegraphics[scale=0.30,angle=0]{625.5}
\includegraphics[scale=0.30,angle=0]{625.3}
\includegraphics[scale=0.30,angle=0]{625-last2.ps}
}
\caption{JVLA images of ID 625 at 1.4, 4.5 and 7.5 GHz. The HPBWs and noise
levels are reported in Tab.~\ref{table2}. Levels are respectively: 
2 2.5 3 5 7 10 30 50 100 200 mJy/beam; -0.2 0.2 0.3 0.5 0.7 1 1.5 2 2.5 3 5 7 10 30 50 70 mJy/beam; 1.5 2 2.5 3 5 7 10 15 20 25 mJy/beam}
\label{625}
\end{figure*}

\noindent
{\bf ID 625}: this source shows at low resolution a central emission
and a double structure in the NS direction (Fig.~\ref{625}) on a scale
of 36\arcsec, $\sim$40 kpc. Toward the South only a diffuse lobe is
visible, while toward the North there is a bright jet-like structure,
typical of FR~Is, terminating into a diffuse plume. This radio
structure is typical of the so-called hybrid radio-sources
\citep{gopal00}. At high resolution the nuclear source is resolved and
rather complex. A compact flat spectrum region (located at 01$^{\rm
  h}$ 48$^\prime$ 16$\farcs$25 00$^{\rm d}$ 19$^\prime$ 44$\farcs$9)
is identified as the radio core based on its flat spectral index (0.05
between 4.5 and 7.5 GHz). From the core a one-sided emission is
visible, pointing at SW, but it sharply bends toward the NE after
$\sim$ 2\arcsec\ from the core and it then enters the N lobe. Two
knots are visible in the 7.5 GHz image with spectral index 0.6 and
1.16 respectively. The large jet bending and the jet asymmetry suggest
that in this source is present a relativistic jet at a small angle
with respect to the line of sight.

\begin{acknowledgements}

RDB was supported at the Technion by a fellowship from the the Lady
Davis Foundation. We thank the anonymous referee for the constructive
comments that have helped us to improve the paper.

\end{acknowledgements}

\bibliography{my.bib}
\end{document}